\documentclass[
reprint,
superscriptaddress,
amsmath,
amssymb,
aps,
longbibliography,
prb,
showpacs,
floatfix
]{revtex4-2}
\usepackage{graphicx} 
\usepackage[dvipsnames]{xcolor}
\usepackage[bookmarks=false,linkcolor=Orange,urlcolor=MidnightBlue,colorlinks,citecolor=Maroon]{hyperref}
\usepackage{physics}
\usepackage{comment}
\usepackage[normalem]{ulem}

\begin{document}
	\title{OAM-mode sorting with a wavefront twister}
		\author{Suman Karan}
	\thanks{These authors contributed equally to this work.}
	\affiliation{Department of Physics, Indian Institute of Technology Kanpur, Kanpur, UP 208016, India}
	\author{Swati Chaudhary}
	\thanks{These authors contributed equally to this work.}
	\affiliation{The Institute for Solid State Physics, The University of Tokyo, Kashiwa, Chiba, Japan}
	\author{Harshwardhan Wanare}
	\affiliation{Department of Physics, Indian Institute of Technology Kanpur, Kanpur, UP 208016, India}
	\author{Anand K. Jha}
	\email{akjha@iitk.ac.in}
	\affiliation{Department of Physics, Indian Institute of Technology Kanpur, Kanpur, UP 208016, India}
	
\begin{abstract}
We propose an OAM sorter based on a novel optical element that we refer to as a wavefront twister. It is a generalization of the conventional wavefront rotators such as the Dove prism. However, unlike a Dove prism, which simply rotates a wavefront, the rotation generated by a wavefront twister varies linearly with radial position, resulting in the twisting of the wavefront. We demonstrate that the wavefront twister, followed by a lens, maps each OAM mode to an annulus of distinct radius at the back focal plane of the lens with negligible inter-modal overlap and preserves the circular symmetry. Thus, the proposed wavefront twister offers a scalable scheme for high-dimensional OAM mode sorting, with important consequences for the practical realization of OAM-based applications.
\end{abstract}

\maketitle
	
\section{Introduction}

Light beams can carry both spin and orbital angular momentum (OAM), associated with circular polarization and helical wavefront respectively \cite{allen1992pra}. Laguerre-Gaussian (LG) beams $LG^{l}_p\left(\rho,\phi\right)$ within the paraxial approximation carry a well-defined OAM of $l \hbar$ per photon, where $l$ is an unbounded integer characterizing the azimuthal phase dependence of the form $e^{i l \phi}$ \cite{allen1992pra}, and $p$ is the radial mode index characterizing the radial profile. Unlike polarization degree of freedom which is limited to two dimensions, the unbounded nature of $l$ makes LG modes a natural high-dimensional basis. Furthermore, Mair et al. \cite{mair2001nature} demonstrated that OAM of photon pairs generated in spontaneous parametric down-conversion is entangled, showing that LG modes can serve as a high-dimensional basis in quantum domain as well. These properties of OAM have led to a wide range of both classical and quantum information applications including high-capacity optical communication through free-space \cite{wang2012natphoton} and optical fiber \cite{bozinovic2013science}, quantum key distribution \cite{cerf2002prl, karimipour2002pra}, quantum gate implementations \cite{ralph2007pra, lanyon2008natphy}, supersensitive angular measurements \cite{jha2011pra}, and for fundamental tests of quantum mechanics \cite{kaszlikowski2000prl, collins2002prl}. 

Harnessing the full capacity of OAM in these applications requires not just detecting OAM modes but sorting them. An OAM detector measures the modal probability distribution \cite{mair2001nature, leach2002prl, kulkarni2017natcomm, karan2025sciadv}. However, a sorter maps each mode to a distinct spatial location, enabling parallel detection, multiplexing, and mode-selective operations. For example, a prism spatially separates different frequency components. A lens maps different transverse wavevectors to distinct positions in the focal plane. A device that performs the analogous operation for OAM modes, an OAM sorter, remains a longstanding challenge.

Existing OAM sorting approaches include interferometric methods based on cascaded Mach-Zehnder interferometers with Dove prisms \cite{leach2002prl, leach2004prl}, which are in principle $100\%$ efficient but require $N-1$ interferometers to sort $N$ modes, making them impractical for large mode sets. Log-polar coordinate transformation based sorters were first demonstrated by Berkhout et al. \cite{berkhout2010prl} using two planar phase elements, sorting up to $l= \pm 5$ but with approximately $70\%$ diffraction loss due to the limited efficiency of the SLM. Lavery et al. \cite{lavery2012oe, lavery2013njp} later improved this efficiency using custom refractive elements, reducing loss to $\approx 15\%$ and extending the range to $l= \pm 25$ at the single-photon level. However, these sorters convert OAM modes into rectangular stripes, wash out the azimuthal phase, and suffer from around $20\%$ crosstalk between consecutive OAM modes due to the finite size of the transformed beam. Mirhosseini et al. \cite{mirhosseini2013natcom} reduced this crosstalk to $2.6\%$ by incorporating  an additional refractive beam-copying element with log-polar transformation \cite{malik2014natcom}. Although the inverse of the log-polar transformation has been demonstrated for OAM multiplexing \cite{huang2015screp}, it has not been demonstrated for the modified technique by Mirhosseini et al. \cite{mirhosseini2013natcom}. In a different approach, the angular lens proposed by Sahu et al. \cite{sahu2018oe} sorts OAM modes into localized spots at separated angular positions using a single phase-only element. However, for experimentally feasible aperture sizes the sorting is possible with a minimum mode separation $\Delta l =\pm 3$. This sorter has also been demonstrated using metamaterials \cite{guo2021lsa}. More recently, a multi-plane light conversion based LG mode sorter has been proposed, but it requires spatially separated input modes to begin with, limiting its applicability as an OAM sorter \cite{fontaine2019natcom}.


Thus, efficient and unambiguous OAM sorting is still an open challenge. To this end, in this article, we propose a novel optical element, which we refer to as a ``wavefront twister.'' Based on this element, we demonstrate an OAM sorting scheme that ensures negligible inter-modal overlap, preserving the circular symmetry of each mode, and scalable to an arbitrarily large OAM mode set.

\section{Proposed Scheme}

\begin{figure}[!t]
\centering
\includegraphics[scale=0.85]{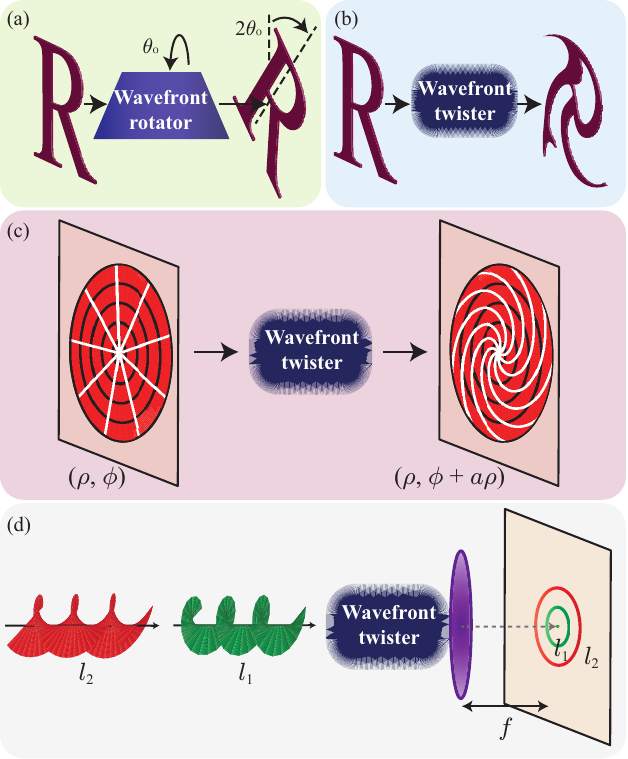}
\caption{Conceptual illustration of the proposed wavefront twister and the OAM sorting scheme. (a) A conventional wavefront rotator such as a Dove prism, oriented at angle $\theta_0$ rotates an input wavefront by angle $2\theta_0$. (b) The proposed wavefront twister rotates an incoming wavefront by an angle $\theta(\rho)=a\rho$ that varies linearly with radial location $\rho$. Consequently, the wavefront is not simply rotated but twisted. (c) The wavefront twister maps the point $(\rho,\phi)$ in the incident wavefront  to $(\rho,\phi + a\rho)$. (d) The proposed OAM sorter scheme consisting of the wavefront twister followed by a lens of focal length $f$ maps the OAM modes $l_1$ and $l_2$ to annuli of distinct radii in the focal plane, leading to their spatial separation.}
\label{fig:conceptualfigure}
\end{figure}

\noindent{\bf The concept of a wavefront twister}\\
Figure \ref{fig:conceptualfigure}(a) illustrates the action of a conventional wavefront rotators such as the Dove prism. A wavefront rotator oriented at angle $\theta_0$ rotates an incoming wavefront by an angle $2\theta_0$. The rotation thus generated does not depend on the details of incoming wavefront. Therefore, in terms of cylindrical coordinates $\left(\rho, \phi, z \right)$, a wavefront rotator maps a point $\left(\rho,\phi\right)$ in the incoming wavefront to the point $\left(\rho,\phi+2\theta_0\right)$ in the outgoing wavefront. In this work, we propose a wavefront twister by generalizing the idea of wavefront rotation. Unlike a wavefront rotator, which simply rotates a wavefront, the rotation generated by a wavefront twister varies linearly with radial position, resulting in the twisting of the incoming wavefront as illustrated in Figs.~\ref{fig:conceptualfigure}(b) and \ref{fig:conceptualfigure}(c). The rotation $\theta (\rho)$ experienced by the incoming wavefront thus depends on the radial position and can be expressed as $\theta\left( \rho\right) = a \rho$, where $a$ characterizes the strength of the twisting. In terms of the cylindrical coordinates, the action of a wavefront twister is to map a point $\left(\rho,\phi\right)$ in the incoming wavefront to the point $\left(\rho,\phi+a\rho\right)$ in the outgoing wavefront, as shown in Fig.~\ref{fig:conceptualfigure}(c).


\

\noindent{\bf The action of a wavefront twister on an LG mode}\\
For Laguerre Gaussian (LG) beams with radial index $p=0$, the incident electric field profile at $z=0$ is given by
\begin{align}\label{eqn:lg_mode}
E_l^{LG}(\rho ,\phi; z=0) &\equiv E_l^{LG}(\rho ,\phi) = E_l^{LG}(\rho)e^{- i l \phi}  \nonumber \\
&= \sqrt{\dfrac{2}{\pi w_0^2 |l|!}} \left( \frac{ \sqrt{2}\rho}{w_0} \right)^{|l|}e^{-\frac{\rho^2}{w_0^2}}e^{- i l \phi},
\end{align}
where $w_0$ is the beam waist of the mode and $l$ is the OAM mode index. Since a wavefront rotator oriented at angle $\theta_0$ rotates an incoming wavefront by $2\theta_0$, an incoming LG mode $ E_l^{LG}(\rho ,\phi)$ given by Eq.~(\ref{eqn:lg_mode}) upon passing through an image rotator gets transformed as
\begin{equation}\label{eqn:doveprism_eqn}
E_l^{LG}(\rho)e^{- i l \phi} \rightarrow E_l^{LG}(\rho)e^{- i l \left(\phi + 2\theta_0\right)}.
\end{equation}
We note that the action of a wavefront rotator on an LG mode of OAM mode index $l$ can be modelled as a linear transformation in which the mode acquires a phase term $e^{-il2\theta_0}$. In an analogous manner, since the rotation experienced by an incoming wavefront passing through a wavefront twister is given by $\theta\left( \rho\right) = a \rho$, the effect of a wavefront twister on an LG mode can be modelled as a linear transformation in which the mode acquires a phase term $e^{-ila\rho}$. This can be expressed as 
\begin{equation}\label{eqn:twister_eqn}
E_l^{LG}(\rho)e^{- i l \phi} \rightarrow  E_l^{LG}(\rho ) e^{-i l \left( \phi + a \rho\right)}.
\end{equation}
We note that the LG modes acquire $l$-dependent radial phase gradients upon passing through a wavefront twister. We show in the next subsection that this feature enables OAM-mode sorting with almost negligible inter-modal overlap.


\

\begin{figure*}[!t]
\centering
\includegraphics[scale=1]{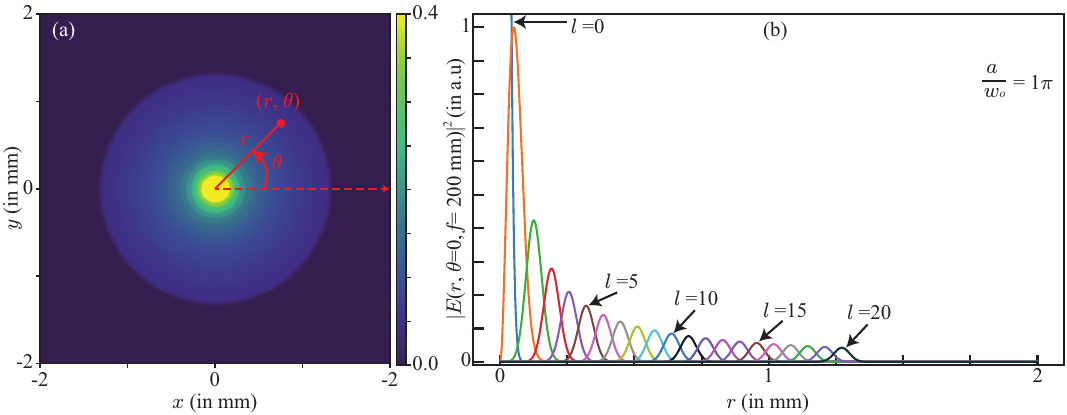}
\caption{(a) Numerically computed two-dimensional intensity $|E(r,\theta, f=200 \ {\rm mm})|^2$ for OAM mode index $l$ ranging from $1$ to $20$ at $a/w_0=1\pi$ and $w_0=0.1$ cm. (b) Numerically computer one-dimensional intensity profiles $|E(r,\theta=0, f=200 \ {\rm mm})|^2$ as a function of $r$ for OAM mode index $l$ ranging from $1$ to $20$ at $a/w_0=1\pi$ and $w_0=0.1$ cm.}
\label{fig:sorted_oam_a1pi}
\end{figure*}
\begin{figure*}[!t]
\centering
\includegraphics[scale=1]{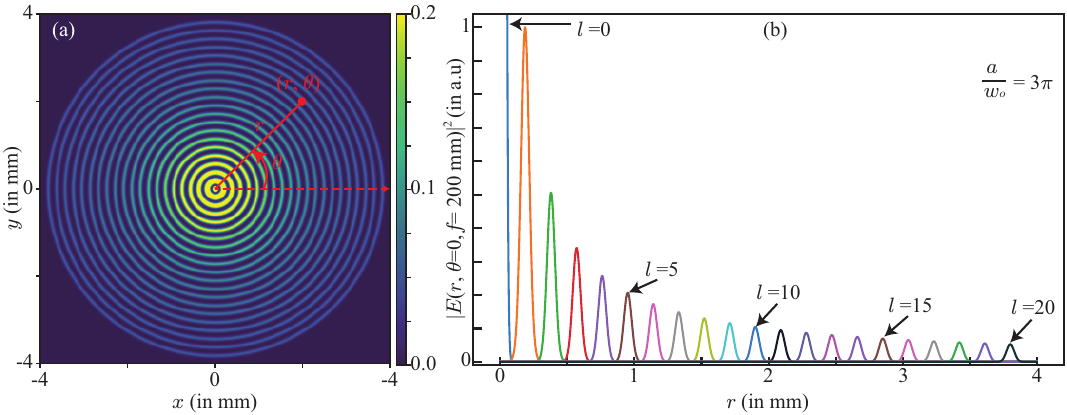}
\caption{(a) Numerically computed two-dimensional intensity $|E(r,\theta, f=200 \ {\rm mm})|^2$ for OAM mode index $l$ ranging from $1$ to $20$ at $a/w_0=3\pi$ and $w_0=0.1$ cm. (b) Numerically computer one-dimensional intensity profiles $|E(r,\theta=0, f=200 \ {\rm mm})|^2$ as a function of $r$ for OAM mode index $l$ ranging from $1$ to $20$ at $a/w_0=3\pi$ and $w_0=0.1$ cm.}
\label{fig:sorted_oam_a3pi}
\end{figure*}

\noindent{\bf OAM-mode sorting with a wavefront twister}\\
Our proposed scheme for OAM-mode sorting is shown in Fig.~\ref{fig:conceptualfigure}(d). A convex lens of focal length $f$ is placed immediately after a wavefront twister and the resulting field distribution is measured at a screen placed at the back focal plane of the lens. For an incident LG mode field, the electric field at the lens, as shown in Fig.~\ref{fig:conceptualfigure}(d), is given by  $ E_l^{LG}(\rho ,\phi) e^{-i l a \rho}$ [Eq.~(\ref{eqn:twister_eqn})]. Using the Fresnel--Kirchhoff diffraction integral, the electric field at a distance $z$ from the lens, $E(r,\theta;z)$ can be written as:
\begin{align}\label{eqn:fresnel_kirchhoff}
 E(r,\theta;z) = & A(k,z)\int^{\infty}_{0}\int^{2\pi}_{0}  E_l^{LG}(\rho,\phi ) e^{-i l a \rho} e^{-i \frac{k}{2f} \rho^2} \nonumber \\
&\times  e^{i \frac{k}{2z}\left[ r^2 - 2r \rho \cos \left(\phi -\theta \right) + \rho^2 \right]} \rho d \rho d\phi.
\end{align}
Here, $(r, \theta)$ represents the cylindrical coordinates at $z$. At $z=f$, the above equation reduces to
\begin{align}\label{eqn:field_at_f_plane}
& E(r,\theta, f) = A(k,z) e^{i \frac{k}{2f}r^2} \nonumber \\
&\times \int_0^{\infty} \int_0^{2\pi} E_l^{LG}(\rho,\phi) e^{-i l a \rho} e^{-i \frac{k r \rho}{f} \cos(\phi - \theta)}\rho d\rho d\phi.
\end{align}
\begin{figure*}[!t]
\centering
\includegraphics[scale=1]{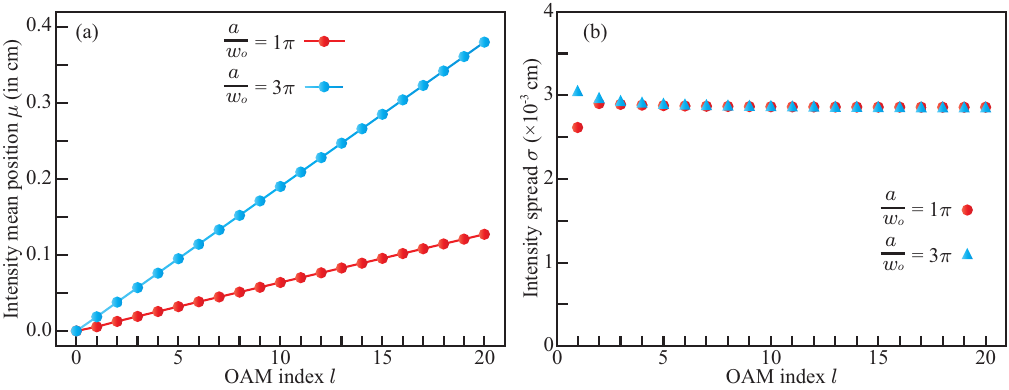}
\caption{(a) Plot of the mean radial position $\mu$ of the ring formed by each sorted OAM mode as a function of the OAM index $l$  (b) Radial spread $\sigma$ of each ring as a function of $l$.}
\label{fig:l_vs_sepration}
\end{figure*}
Substituting from Eq.~(\ref{eqn:lg_mode}) and using the identity $e^{-i \frac{k r \rho}{f} \cos(\phi -\theta)}=\sum_n \left(-i\right)^n J_n\left(\frac{k r \rho}{f}\right) e^{- i n\theta}e^{i n\phi}$, Eq.~(\ref{eqn:field_at_f_plane}) reduces to a one-dimensional integral:
\begin{align}\label{eqn:oned_rho_integral}
E(r,\theta, f) = 2\pi \sqrt{\dfrac{2}{\pi w_0^2 |l|!}} \left(-i\right)^l A(k,z) e^{i \frac{k}{2f}r^2} e^{- i l \theta}\nonumber \\
\times \int_0^{\infty}  e^{-\frac{\rho^2}{w_0^2}}  e^{-i l a \rho} J_l\left(\frac{k r \rho}{f}\right)\left( \frac{ \sqrt{2}\rho}{w_0} \right)^{|l|} \rho d\rho,
\end{align}
which can also be evaluated as an infinite summation of hypergeometric functions (see Appendix ~\ref{app:with_twister_calculation} for a detailed derivation).  We numerically compute $|E(r,\theta, z=f)|^2$  from the integral  in Eq.~(\ref{eqn:oned_rho_integral}) for different OAM mode index $l$ and plot the results in Figs.~\ref{fig:sorted_oam_a1pi} and \ref{fig:sorted_oam_a3pi}.


\section{Results}
\begin{figure}[!b]
\centering
\includegraphics[scale=0.96]{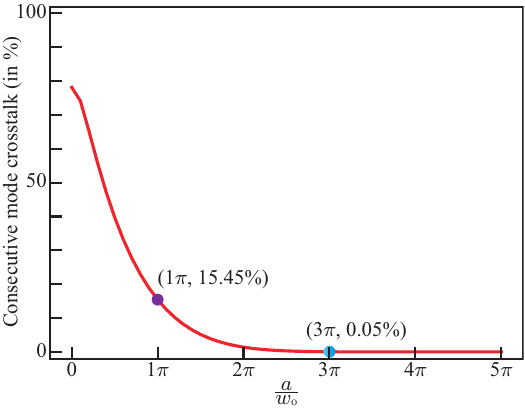}
\caption{Plot of consecutive mode crosstalk as a function of  $a/w_0$. The overlap percentage decreases monotonically with increasing $a/w_0$. The highlighted points correspond to crosstalk percentages for $a/w_0 = 1\pi$ and $3\pi$.}
\label{fig:overlap_vs_aa}
\end{figure}
Fig.~\ref{fig:sorted_oam_a1pi} (a) shows the numerically computed two-dimensional intensity profiles $|E(r,\theta, f=200 \ {\rm mm})|^2$ for different $l$ values at $a/w_0 = 1\pi$ and $w_0 =0.1$ cm. Fig.~\ref{fig:sorted_oam_a1pi} (b) plots the one-dimensional intensity profiles $|E(r,\theta=0, f=200 \ {\rm mm})|^2$ as a function of $r$ for different $l$ values at $a/w_0 = 1\pi$ and $w_0 =0.1$ cm. We find that the overlap between adjacent $l$ modes decreases significantly compared to the unsorted LG profile. For $a/w_0 = 1\pi$, the separation between adjacent $l$ states exceeds $\sigma$, resulting in a consecutive mode crosstalk of less than $15.45\%$. Increasing $a/w_0$ to $3\pi$ reduces the crosstalk to about $0.05\%$, as shown in Fig.~\ref{fig:sorted_oam_a3pi} (a) and \ref{fig:sorted_oam_a3pi} (b). The radial intensity profiles of each sorted mode is well approximated by a Gaussian. Fig.~\ref{fig:l_vs_sepration}(a) shows the mean radial position $\mu$ as a function of $l$ for both $a/w_0=1\pi$ and $a/w_0=3\pi$, confirming that $\mu$ increases linearly with $l$ and that the separation between consecutive modes increases with $a/w_0$. Fig.~\ref{fig:l_vs_sepration}(b) shows the standard deviation $\sigma$ for the fitted Gaussian which we refer to as intensity spread as a function of $l$ for the same two values of $a/w_0$. The spread $\sigma$ remains nearly constant across all $l$ values and does not change with $a/w_0$, meaning the sorting efficiency depends only on $a/w_0$ and not on the mode index. Thus, in principle, this scheme can sort an arbitrarily large number of modes with the same efficiency.   

In order to analyze the sorting capability as a function of the twisting strength, we plot the consecutive OAM mode crosstalk percentage as a function of $a/w_0$ in Fig.~\ref{fig:overlap_vs_aa}. The crosstalk decreases monotonically with increasing $a/w_0$. The two highlighted points correspond to $a/w_0 = 1 \pi$ and $3\pi$, for which the results are shown in Figs.~\ref{fig:sorted_oam_a1pi} and \ref{fig:sorted_oam_a3pi}. The crosstalk percentages at these two values are $15.45\%$ and $0.05\%$, respectively, confirming that the sorting efficiency improves continuously with the twisting strength.	

\section{Conclusions and Discussion}
We have proposed a wavefront twister, an optical element that twist an incoming wavefront such that the rotation experienced by the wavefront varies linearly with radial position. Utilizing this optical element, we have proposed an OAM-mode sorting scheme. The numerical results show that our scheme sorts OAM modes with negligible inter-modal overlap. We note that the proposed OAM sorting scheme cannot distinguish between $l$ and $-l$ OAM mode indices, as both produce annular intensity distributions at the same radial position. Furthermore, the coordinate transformation $\left(\rho,\phi\right) \rightarrow \left(\rho ,\phi+ a\rho \right)$ cannot in general be implemented by a single phase element. This is true even for realizing a wavefront rotator in the form of a Dove prism or K-mirrors, for which three distinct planar phase elements are required including one out-of-plane reflector. Our proposed scheme is scalable to arbitrarily large OAM mode sets. Therefore, an experimental realization of a wavefront twister can have several important consequences for the practical realization of OAM-based quantum and classical communication schemes.

\section{Acknowledgements}

We acknowledge financial support from the Science and Engineering Research Board through grants STR/2021/000035  \& CRG/2022/003070, and from the Department of Science \& Technology, Government of India through grant DST/ICPS/QuST/Theme-I/2019 and through the National Quantum Mission (NQM) technical group (TG) project on quantum imaging.

\bibliographystyle{abbrv}
\bibliography{twister_based_oam_sorter_ref}

\onecolumngrid
\appendix
\section{Field distribution in the back focal plane of the lens with the wavefront twister}\label{app:with_twister_calculation}
We evaluate the radial integral in Eq.~(\ref{eqn:oned_rho_integral}), which is given by
\begin{align}\label{eqn:only_rho_integral}
A_l\left(r\right)=  \int_0^{\infty}  e^{-\frac{\rho^2}{w_0^2}}  e^{-i l a \rho} J_l\left(\frac{k r \rho}{f}\right)\left( \frac{ \sqrt{2}\rho}{w_0} \right)^{|l|} \rho d\rho. 
\end{align}
To evaluate this integral, we first expand  $e^{-i l a \rho}$ as a power series of the form
\begin{equation} \label{eqn:exp_power_series}
e^{-ila \rho} = \sum^{\infty}_{m=0} \frac{\left(-i l a \rho \right)^m}{m!} = \sum^{\infty}_{m=0} \frac{\left(-i l a \right)^m}{m!} \rho^m.
\end{equation}
Next, substituting this expansion and using the identity $J_l\left(\frac{k r \rho}{f}\right)= (-1)^{\frac{|l|-l}{2}}J_{|l|}\left(\frac{k r \rho}{f}\right)$ in Eq.~(\ref{eqn:only_rho_integral}), we get
\begin{equation}\label{eqn:alr_expansion}
A_l\left(r\right) = \sum^{\infty}_{m=0} \frac{\left(-i l a \right)^m}{m!}\left(\frac{\sqrt{2}}{w_0}\right)^{|l|}\left(-1\right)^{\frac{|l| -l}{2}} \int_0^{\infty}  e^{-\frac{\rho^2}{w_0^2}}  J_{|l|}\left(\frac{k r \rho}{f}\right) \rho^{|l| +m +1} d\rho. 
\end{equation}
The integral in each term of the summation can be evaluated using the identity from Gradshteyn \& Ryzhik (6.631-1)~\cite{gradshteyn2014}:
\begin{equation}\label{eqn:integration_identity}
\int^{\infty}_{0} x^{\mu} e^{-\alpha x^2} J_{\nu}\left(\beta x\right) dx = \frac{\beta^{\nu} \Gamma\left(\frac{\nu +\mu +1}{2}\right)}{2^{\nu +1} \alpha^{\frac{\nu +\mu +1}{2}} \Gamma\left(\nu +1\right)} {}_1F_1\left[\frac{\nu +\mu +1}{2};\nu +1; - \frac{\beta^2}{4 \alpha}\right].
\end{equation}
Setting $\alpha= 1/w^2_0$, $\mu = |l| +m +1$, $\nu =|l|$, and $\beta = k r /f$ in the above identity, Eq.~(\ref{eqn:alr_expansion}) reduces to
\begin{equation}
A_l\left(r\right) = \sum^{\infty}_{m=0} \frac{\left(-i l a \right)^m}{m!}\left(\frac{\sqrt{2}k r}{w_0 f}\right)^{|l|}\left(-1\right)^{\frac{|l| -l}{2}}\frac{ \Gamma\left(|l| +1 + \frac{m}{2}\right) \left(w^2_0\right)^{|l| + 1 + \frac{m}{2}}}{2^{|l|+1}  \Gamma\left(|l| +1\right)} {}_1F_1\left[|l|+1+ \frac{m}{2};|l| +1; - \frac{k^2 r^2 w^2_0}{4 f^2}\right] .
\end{equation}
$A_l\left(r\right)$ is thus an infinite summation of hypergeometric functions ${}_1F_1$.  We evaluate the integral in Eq.~(\ref{eqn:oned_rho_integral}) numerically.

\end{document}